\documentclass[prm,twocolumn,
citeautoscript,superscriptaddress]{revtex4-1}

\setcitestyle{super}

\usepackage{hyperref}

\hypersetup{colorlinks=true,breaklinks=true,
	urlcolor=blue,linkcolor=blue,
	bookmarksopen=true,
	citecolor=blue}

\usepackage[T1]{fontenc}\usepackage[latin1]{inputenc}

\usepackage{amsmath}
\usepackage{times}
\usepackage{amssymb}
\usepackage{bbold} 
\usepackage{graphicx}
\usepackage{dcolumn,graphicx,color,booktabs,microtype,afterpage}

\usepackage{bm}
\usepackage{braket}

\usepackage{times}

\usepackage{MnSymbol}

\usepackage{color}
\usepackage[normalem]{ulem}

\begin{document}

\title{Imprinting and driving electronic orbital magnetism using magnons}

\author{Li-chuan~Zhang}
\affiliation{Peter Gr\"unberg Institut and Institute for Advanced Simulation, Forschungszentrum J\"ulich and JARA, 52425 J\"ulich, Germany}
\affiliation{Department of Physics, RWTH Aachen University, 52056 Aachen, Germany}
\author{Dongwook Go}
\affiliation{Peter Gr\"unberg Institut and Institute for Advanced Simulation, Forschungszentrum J\"ulich and JARA, 52425 J\"ulich, Germany}
\affiliation{Institute of Physics, Johannes Gutenberg University Mainz, 55099 Mainz, Germany}
\author{Jan-Philipp~Hanke}
\affiliation{Peter Gr\"unberg Institut and Institute for Advanced Simulation, Forschungszentrum J\"ulich and JARA, 52425 J\"ulich, Germany}
\author{Patrick~M.~Buhl}
\affiliation{Institute of Physics, Johannes Gutenberg University Mainz, 55099 Mainz, Germany}
\author{Sergii~Grytsiuk}
\affiliation{Peter Gr\"unberg Institut and Institute for Advanced Simulation, Forschungszentrum J\"ulich and JARA, 52425 J\"ulich, Germany}
\author{Stefan~Bl\"ugel}
\affiliation{Peter Gr\"unberg Institut and Institute for Advanced Simulation, Forschungszentrum J\"ulich and JARA, 52425 J\"ulich, Germany}
\author{Fabian~R.~Lux}
\affiliation{Peter Gr\"unberg Institut and Institute for Advanced Simulation, Forschungszentrum J\"ulich and JARA, 52425 J\"ulich, Germany}
\affiliation{Department of Physics, RWTH Aachen University, 52056 Aachen, Germany}
\author{Yuriy~Mokrousov}\email[Corresponding author:~]{y.mokrousov@fz-juelich.de}
\affiliation{Peter Gr\"unberg Institut and Institute for Advanced Simulation, Forschungszentrum J\"ulich and JARA, 52425 J\"ulich, Germany}
\affiliation{Institute of Physics, Johannes Gutenberg University Mainz, 55099 Mainz, Germany}

\begin{abstract}
\noindent
{\large{\bf Abstract}}\\
Magnons, as the most elementary excitations of magnetic materials, have recently emerged as a prominent tool in electrical and thermal manipulation and transport of spin, and magnonics as a field is considered as one of the pillars of modern spintronics. On the other hand, orbitronics, which exploits the orbital degree of freedom of electrons rather than their spin, emerges as a powerful platform in efficient design of currents and redistribution of  angular momentum  in structurally complex materials.  
Here, we uncover a way to bridge the worlds of magnonics and electronic orbital magnetism, which originates in the fundamental coupling of scalar spin chirality, inherent to magnons, to the orbital degree of freedom in solids. We show that this can result in efficient generation and transport of electronic orbital angular momentum by magnons, thus opening the road to combining the functionalities of magnonics and orbitronics to their mutual benefit in the realm of spintronics applications. 
\end{abstract}

\maketitle

\noindent
{\large{\bf Introduction}}\\
Spin-heat conversion is a guiding motive in spin caloritronics, which sets out to explore physical phenomena beyond the limits of conventional electronics for energy-efficient information processing~\cite{vzutic2004spintronics,bauer2012spin,boona2014spin,chumak2015magnon}. In this light, spin-wave excitations, 
known as magnons, offer bright prospects as they mediate thermal spin transport via analogs of 
Seebeck~\cite{geballe1955seebeck,uchida2008observation,xiao2010theory} and Nernst effects in magnetic and non-magnetic materials~\cite{kikkawa2013longitudinal,miyasato2007crossover,kovalev2016spin,meyer2017observation}. 
It has been suggested that the complex spin arrangement exhibited by antiferromagnets and non-collinear 
magnets provides an alternative route for triggering spin-heat conversion through magnons~\cite{menzel2012information}, which relies on a chiral  coupling 
$\mathbf{S}_i \times \mathbf{S}_j$ among spin moments $\mathbf{S}_i$ and $\mathbf{S}_j$.
However, while converting temperature gradients into transverse spin currents as a consequence of the Dzyaloshinskii-Moriya interaction~\cite{sergienko2006role,heide2008dzyaloshinskii,zhang2013topological, cheng2016spin,kovalev2016spin}, the spin Nernst effect of magnons is rather inefficient in light materials as it is  proportional to the strength of relativistic spin-orbit interaction~\cite{cheng2016spin,kovalev2016spin}. 

The situation here is quite similar to the one we are facing in the realm of the spin Hall effect~\cite{hirsch1999spin}, where strong spin-orbit interaction is prerogative for generation of sizeable spin currents which can be in turn used to e.g. switch the magnetization via the effect of spin-orbit torque~\cite{miron2010current,liu2012spin,garello2013symmetry}. Recently, a new paradigm has emerged which relies on the generation of the currents of orbital angular momentum rather than spin, and which carries many advantages over conventional protocols in spinorbitronics~\cite{kontani2009giant,tanaka2008intrinsic,go2018intrinsic,go2020orbital}. The corresponding palette of effects evolving around  orbitronics is largely grounded in the fundamental fact that the electric field driven currents of orbital angular momentum of electrons as a rule overshadow the accompanying currents of spin by far, while remaining large even in the lightest materials~\cite{kontani2009giant,go2018intrinsic}. Although nothing is known about interaction of magnonic excitations with electronic orbital magnetism, it appears to be extremely beneficial to marry the promising ideas of orbitronics with the magnon-based philosophy which has been very successful in the domain of spin transport and spin caloritronics so far.

\begin{figure*}
\centering
\includegraphics [scale=0.25,trim=0 0 80 0]{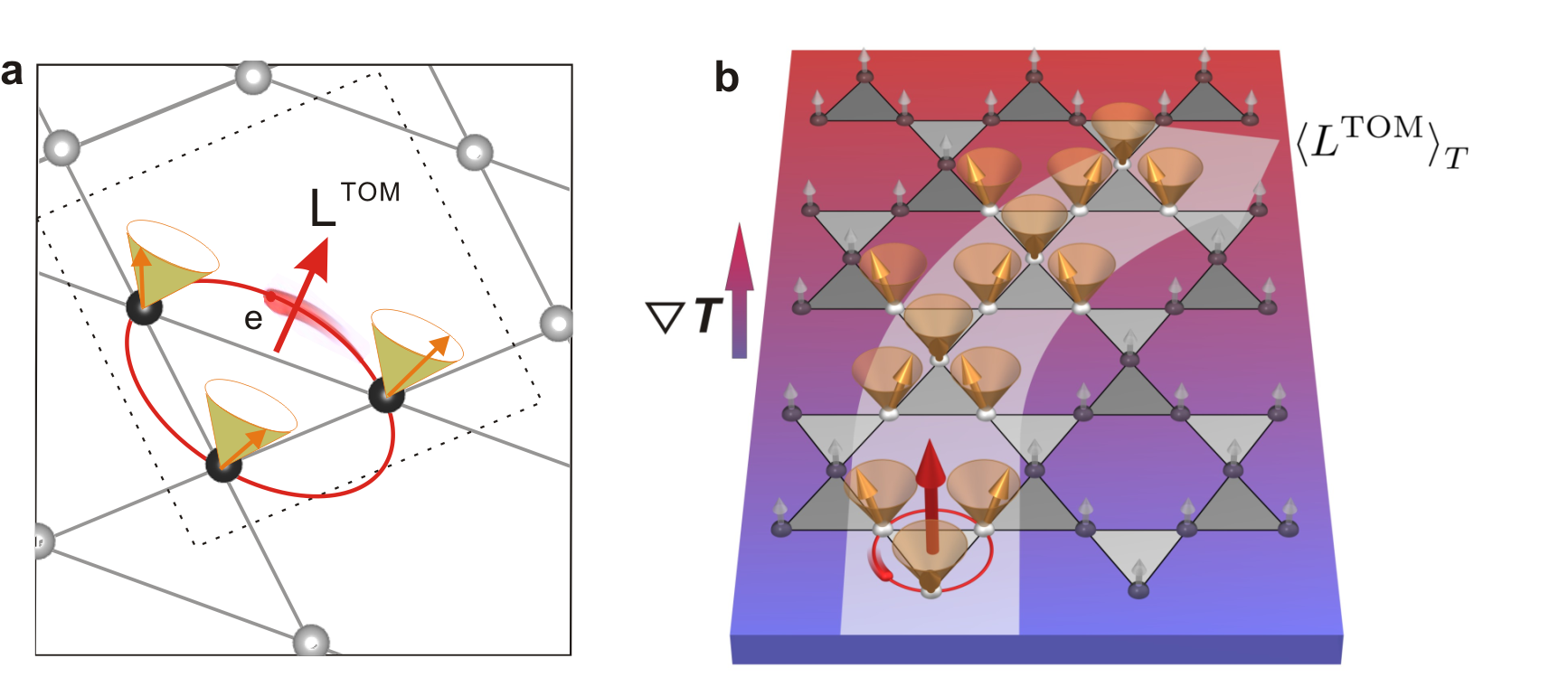} 
\caption{{\bf Generation and drag of orbital angular momentum by magnons}. {\bf a}~An electron hopping among non-collinear triplets of spins gives rise to so-called topological orbital moment (TOM), $\mathbf{L}^{\rm TOM}$, which points out of the plane of the spins. The electronic TOM is effectively induced by the scalar spin chirality realized for example on a kagome spin lattice, which is shown in an oblique view. The unit cell is outlined with the dotted line. {\bf b}~Sketch of the orbital Nernst effect of magnons for a ferromagnet on an example  kagome lattice. While the generation of a magnon (orange arrows) imprints an average scalar spin chirality into the system and leads to the generation of electronic TOM (red arrow), the generation of a magnon flow in a temperature gradient $\nabla T$  results in a transverse deflection of magnons and corresponding  TOM-mediated drag of the orbital angular momentum, denoted by $\braket{ L^\mathrm{TOM}}$ $-$ which we refer to as the orbital Nernst effect. 
}
\label{fig1}
\end{figure*}

Nowadays, various magnetic phenomena in chiral spin systems are often interpreted based on a second flavor of chirality $-$ the scalar spin chirality (SSC) $\chi_{ijk}=\hat{\mathbf{S}}_i \cdot(\hat{\mathbf{S}}_j \times\hat{\mathbf{S}}_k)$ 
between triplets of spins, where $\hat{\mathbf{S}}_{\alpha}$ is the unit vector along $\mathbf{S}_{\alpha}$. 
This type of chirality,  which is inherent to skyrmions~\cite{nagaosa2013topological,seki2012observation,dos2016chirality,lux2018engineering,redies2019distinct} and frustrated magnets~\cite{taguchi2001spin,fujimoto2009hall,diep2013frustrated}, has been crucial for understanding of e.g. topological Hall effect~\cite{neubauer2009topological,kanazawa2011large}.  In the context of skyrmions the SSC is known as the emergent magnetic field which impacts the dynamics of electrons in a way similar to usual but spin-dependent magnetic field~\cite{nagaosa2013topological}.   
In particular, it has been recently realized that the presence of SSC in frustrated magnets and skyrmions reflects in a non-vanishing contribution to the orbital moment of electrons hopping among triplets of non-coplanar spins $-$ just as applying usual magnetic field would give rise to orbital magnetization, see Fig.~\ref{fig1}{\bf a}~\cite{hoffmann2015topological,hanke2016role,lux2018engineering}. 
 The emergence of such chirality-driven orbital magnetization in various systems 
 has been shown in recent years 
 ~\cite{hoffmann2015topological,hanke2016role,hanke2017prototypical,dos2016chirality,lux2018engineering,grytsiuk2020topological,wimmer2019chirality}.  

Here, we explore the idea that the coupling between chirality and electronic orbital magnetism presents a unique way to harvest orbital angular momentum by generating magnons.
We ask the question whether magnonic excitations themselves can give rise to net SSC, even if it is absent in the ground state.
If yes, then generating SSC by magnons would provide a unique mechanism for imprinting electronic orbital angular momentum into the system.
Further, since an applied temperature gradient can drive magnon scattering, it is
reasonable to ask whether this can result in a magnon ``drag'' of orbital angular momentum. If present, such an effect, Fig.~\ref{fig1}{\bf b},
would give an ability of driving  orbital currents by magnons in addition to currents of spin. 
We provide confirmative answers to both questions, and
discuss possible implications of our findings 
for spintronics applications.

\vspace{0.5cm}
\noindent
{\large{\bf Results}}\\
In order to demonstrate the emergence of magnon-mediated orbital magnetization and generation of the current of electronic orbital angular momentum, we refer to microscopic tight-binding and effective spin model of a ferromagnet on a kagome lattice. Conceptually, we separately consider the properties of the electronic bath which exhibits topological orbital magnetism, and the properties of the reservoir of magnons, while coupling both parts of the system by effective topological orbital electron-magnon interaction. First, we present the results which concern the generation of electronic orbital magnetism by the mechanism of SSC. 

\vspace{0.5cm}
\noindent
{\bf Electronic topological orbital magnetism.} We model the electronic part of our system by making use of the tight-binding model of a magnet on a two-dimensional (in the $xy$-plane) kagome lattice (Fig.~1{\bf a}), whose details are explained in Methods. The electronic Hamiltonian is set by considering hoppings among the atoms and an exchange splitting at each atomic site, in a way similar to that of Refs.~\cite{chen2014anomalous, Zhang_2018}. To uncover the SSC-mediated mechanism of orbital moment generation, the spin-orbit interaction is explicitly not taken into account. We start with the ferromagnetic state with the spins pointing out of the plane (see the corresponding band structure in Fig.~2{\bf a}) and then rotate all spins into the plane by an angle $\theta$ away from the $z$-axis, while  keeping the azimuthal angles of the three spins at $0^{\circ}$, $120^{\circ}$ and $240^{\circ}$ (keeping $z$-axis as the three-fold rotational symmetry axis). We find that the effect of such non-coplanarity on the band structure is most prominent in the vicinity of band degeneracies, Fig.~2{\bf a}.

As has been realized in the past years, the non-vanishing SSC in  canted spin systems gives rise to a special type of electronic orbital moment $-$ the topological orbital moment (TOM). While being in its essence a Berry phase effect, microscopically, such TOM arises in response to breaking of symmetry by non-coplanarity, which allows for formation of non-local persistent orbital currents of electrons without any need for spin-orbit interaction~\cite{shindou2001orbital,tatara2002chirality,PhysRevB.78.024402,lux2018engineering,grytsiuk2020topological}. 
 The emergence of topological orbital magnetization in various systems, including celebrated MnGe and Mn$_3$Ge compounds, has been shown in recent years from effective models, tight-binding and first-principles calculations~\cite{hoffmann2015topological,hanke2016role,hanke2017prototypical,dos2016chirality,lux2018engineering,grytsiuk2020topological,wimmer2019chirality}. The Zeeman interaction of TOM with an external magnetic field is known as the ring exchange, which  contributes to the spin Hamiltonian of chiral spin systems~\cite{sen1995large,gritsev2004phase,katsura2010theory}.
By its nature, the TOM can be phenomenologically expressed in terms of the SSC as~\cite{lux2018engineering,grytsiuk2020topological,hanke2017prototypical}:
\begin{eqnarray}
\mathbf{L}^\mathrm{TOM}=\kappa^\mathrm{TO}\sum_{\langle ijk \rangle}  \hat{\mathbf{e}}^{ijk} \chi_{ijk},
\label{eq:TOM}
\end{eqnarray}
where $\langle ijk \rangle$ indicates that spins $i$, $j$, and $k$ are  nearest neighbors  forming a triangle (Fig. 1{\bf a}), and a unit vector $ \hat{\mathbf{e}}^{ijk}$ is normal to the triangle plane. The constant $\kappa^{\rm TO}$ is known as the topological orbital susceptibility~\cite{lux2018engineering,grytsiuk2020topological} and it characterizes the strength of the orbital response of electrons to the SSC. It has been shown that for materials with relatively small spin-orbit strength, the influence of the spin-orbit interaction on TOM is minor~\cite{hanke2017prototypical,grytsiuk2020topological}. This is in contrast to collinear magnets, where the orbital magnetism appears solely as a result of spin-orbit interaction in the system. Here, we do not consider the so-called chiral, proportional to the vector spin chirality, contribution to the orbital magnetism in the system, as it is expected to arise in the regime of large spin-orbit interaction, not considered in this work~\cite{lux2018engineering}.

\begin{figure*}[ht!]
\centering
\includegraphics [scale=0.345,trim=80 0 50 0]{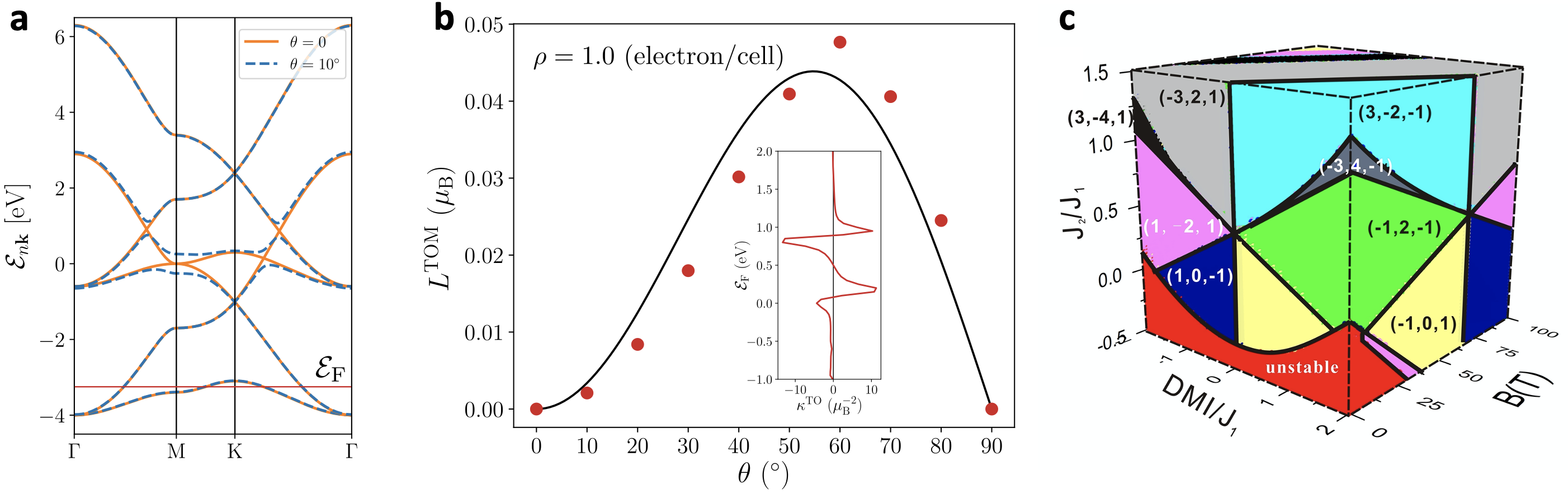}
\caption{{\bf Microscopics of topological orbital magnetism and magnonic topology of the model}. {\bf a}~The electronic band structure based on the tight-binding model of a kagome ferromagnet. The orange lines represent the bands of the  ferromagnetic structure and the blue dotted lines mark the bands of the state canted by a polar angle of $\theta=10^{\circ}$.   {\bf b}~The total toplogical orbital moment (TOM)  as a function of the canting angle for the electron density of  $\rho=1.0$\,$e (cell)^{-1}$. The red symbols mark the calculated values according to Eq.~\eqref{eq:modern}, while the black line is a fit according to  Eq.~\eqref{eq:TOM}. The inset displays the value of the topological orbital susceptibility $\kappa^{\rm TO}$ around the ferromagnetic state as a function of Fermi energy of the tight-binding model. {\bf c}~Topological phase diagram of the magnonic bands of a kagome ferromagnet as a function of the second nearest-neighbor Heisenberg coupling  $J_2$ and Dzyaloshinskii-Moriya interaction (DMI) (in units of the nearest-neighbor Heisenberg coupling $J_1$), as well as external magnetic field $B$ (in Tesla). Colors highlight different phases that are characterized by sets ($C_1$,$C_2$,$C_3$) of Chern numbers. The unstable ferromagnetic phase is shown in red.
}
\label{TOM_and_phase}
\end{figure*}

For our electronic system we numerically access the response of electronic orbital magnetization to canting by referring to the rigorous expression:
\begin{eqnarray}
\mathbf{L}^\mathrm{TOM} 
& &
= 
\frac{\rm{e}}{2 \hbar}
\sum_{n\mathbf{k}\in \mathrm{occ}}
\mathrm{Im}
\left[
\bra{\partial_\mathbf{k} u_{n\mathbf{k}}}
\times 
\right.
\nonumber
\\
& &
\left.
\left\{
\mathcal{H} (\mathbf{k}) + \mathcal{E}_{n\mathbf{k}} - 2\mathcal{E}_\mathrm{F}
\right\}
\ket{\partial_\mathbf{k} u_{n\mathbf{k}}}
\right],
\label{eq:modern}
\end{eqnarray}
where $\mathcal{H}(\mathbf{k})$ is an effective single-particle tight-binding electronic Hamiltonian of our system canted by an angle $\theta$, $u_{n\mathbf{k}}$ is a periodic part of Bloch state with band index $n$ and crystal momentum $\mathbf{k}$, its corresponding energy eigenvalue is $\mathcal{E}_{n\mathbf{k}}$, and the summation goes over all occupied states below the Fermi energy $\mathcal{E}_\mathrm{F}$. We  analyze the behavior of $\mathbf{L}^\mathrm{TOM}$
as a function of angle $\theta$, and compare it to that expected from  Eq.~\ref{eq:TOM}, finding that, overall, the explicitly calculated orbital response of the system to canting fits the TOM-picture very well, see for example the case with band filling of one electron per unit cell in Fig.~2{\bf b}. In accord to this picture, the orbital moment vanishes for the coplanar and collinear cases, and the largest value of TOM is reached for the state with largest SSC. This type of behavior, when $\kappa^{\rm TO}$ with a good degree of accuracy can be assumed to be independent of $\theta$ in the whole range of possible canting, persists over large regions of energies. 
The deviations from it occur in the vicinity of band crossings where the response of the band structure to canting is very large, and where the orbital response is expected to be pronounced~\cite{lux2018engineering}. 

Regardless, in our work we focus on the interplay of orbital magnetism and magnons which cause small deviations of the magnetization from the ferromagnetic state, thus the value of the topological orbital susceptibility in the vicinity of $\theta=0^{\circ}$ is of primary interest. Our calculations, shown in the inset of Fig.~2{\bf b} for the entire range of energies of the model, reveal that the magnitude of $\kappa^{\rm TO}$ in the limit of small canting exceeds the value of 1\,$\mu_\mathrm B$ over large regions of energy, and sensitively depends on the electronic structure.

Overall, our calculations demonstrate that even within the simplest electronic structure considered here it is possible to generate sizable electronic  orbital magnetization by the mechanism of SSC, the properties of which can be tuned by electronic structure design. We show below how to exploit the SSC generation by magnons in order to imprint electronic orbital magnetism into the system.

\begin{figure*}[ht!]
\centering
\includegraphics[scale=0.36,trim=10 20 0 0]{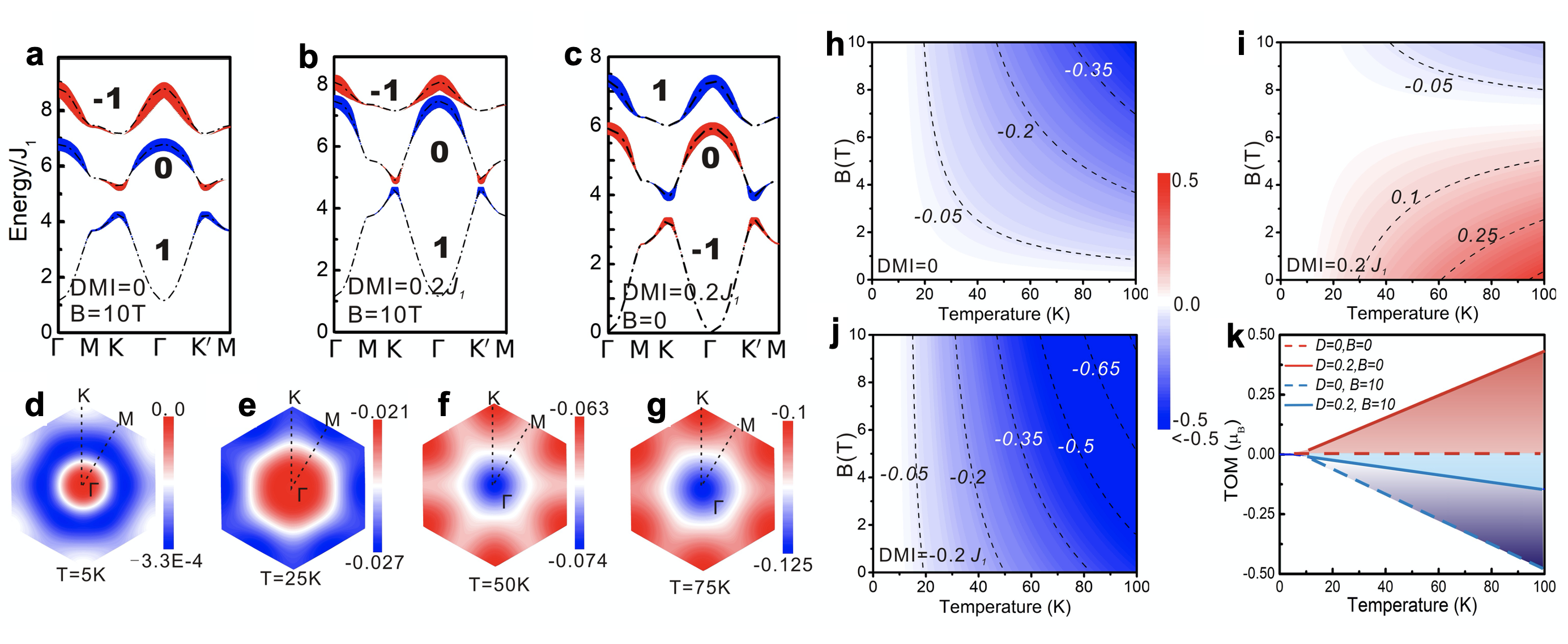}
\caption{{\bf Imprinting electronic orbital magnetism by magnons in a kagome ferromagnet}. {\bf a--c}~Fat band analysis for the magnonic bands of the model for the values of the Dzyaloshinskii-Moriya interaction (DMI) (in units of the nearest-neighbor Heisenberg coupling $J_1$), and magnetic field $B$ specified at the bottom. Red and blue colors represent positive and negative sign of the local topological orbital moment (TOM) $L_{n\mathbf k}^\mathrm{TOM}$, respectively, and the line thickness denotes the corresponding magnitude. Bold integers indicate the Chern numbers of the spin-wave bands. {\bf d--g}~Distribution of the local TOM in the Brillouin zone for different temperatures, after summing over all magnon branches weighted by the Bose distribution. The color map is in units of $\mu_\mathrm{B}$, and the model parameters of panel~{\bf a} are used. {\bf h--k}~Overall TOM of the spin-wave system as a function of magnetic field and temperature. The panels {\bf h--j} present phase diagrams for the DMI strengths of $0$, $0.2J_1$, and $-0.2J_1$, respectively, with the color map indicating the net TOM in units of $\mu_\mathrm{B}$ per unit cell. In~{\bf k}, solid and dotted lines correspond to DMI strengths of $0$ and $0.2J_1$, respectively, and the magnetic field is given in Tesla.}
\label{fig2}
\end{figure*}

\vspace{0.5cm}
\noindent
{\bf  Modelling magnonic excitations.} We consider the effect of magnons on electronic orbital magnetism by referring to an effective Hamiltonian of spin waves of a ferromagnet on a two-dimensional kagome lattice, which is
given by
\begin{equation}
\begin{split}
    H=&-\frac{1}{2}\sum_{ij}J_{ij}  \mathbf{S}_i \cdot \mathbf{S}_j -\frac{1}{2}\sum_{ij}\mathbf{D}_{ij} \cdot (\mathbf{S}_i\times \mathbf{S}_j)\\  &-\mathbf{B}\cdot \kappa^\mathrm{TO}\sum_{ijk} \hat{\mathbf e}_{ijk} 
    [
    \hat{\mathbf{S}}_i \cdot (
    \hat{\mathbf{S}}_j \times
    \hat{\mathbf{S}}_k)] 
    -{{\mu_\mathrm B}} \mathbf{B}\cdot\sum_i \mathbf{S}_i  \, ,
\end{split}
\label{Ham}
\end{equation}
 where $J_{ij}$  mediates the Heisenberg exchange between spins $\mathbf{S}_{i}$ and $\mathbf{S}_{j}$ on sites $i$ and $j$, 
 the second term is the antisymmetric Dzyaloshinskii-Moriya interaction (DMI) quantified by vectors $\mathbf{D}_{ij}$, and the fourth term couples the spins to an external magnetic field $\mathbf B$. In addition, we extend the Hamiltonian by the ring-exchange term in Eq.~\eqref{Ham} to include explicitly the interaction between the magnetic field and the TOM~\cite{PhysRevB.73.155115,sen1995large,katsura2010theory}. This term is given by the product of the SSC  
 and the topological orbital susceptibility $\kappa^\mathrm{TO}$
 ~\cite{lux2018engineering,grytsiuk2020topological}. 
 Owing to the symmetry of the planar kagome lattice, the TOM and the DMI vectors are perpendicular to the film plane (along the $z$-axis), along which we also apply the external magnetic field of magnitude $B$. 
 
 We consider in our analysis only nearest-neighbor interactions except for the Heisenberg term, where we include next-nearest neighbors as well. We set the nearest-neighbor Heisenberg coupling to $J_1=1  meV$, the next-nearest neighbor strength amounts to $J_2=0.1\,J_1$ unless stated otherwise, and the spin-moment length $S$ is fixed to $1$.
 For the magnitude of  topological orbital susceptibility $\kappa^\mathrm{TO}$ we choose a representative value of  $-0.5\,\mu_\mathrm B$ $-$ a value not only motivated by recent material studies~\cite{hanke2016role,grytsiuk2020topological,wimmer2019chirality}, but also corresponding to the lower bound of  $\kappa^\mathrm{TO}$-range found above for small deviations from the ferromagnetic state. 
 As follows from model considerations, the range of values for $\kappa^\mathrm{TO}$ exhibited by the electrons living on a kagome lattice is very large, and one should keep in mind that the effects discussed below can be further enhanced by engineering the electronic structure and the values of $\kappa^\mathrm{TO}$. 
 This route of material design is distinctly different from that associated with the design of the spin-orbit strength,  taken routinely in conventional spinorbitronics.

We first analyze the magnonic bands and their topology in Fig. 2{\bf c} and Fig. 3{\bf a-c}.  The dispersion of the three spin-wave branches in the presence of an external magnetic field of $10$\,T, is presented in Fig. 3{\bf a}. In the absence of DMI, the different magnon bands exhibit Chern numbers $1$, $0$, and $-1$ solely due to the coupling of the magnetic field to the SSC manifesting in a non-zero TOM carried by the magnons, as we show below. 
By including the effect of DMI,  Fig. 3{\bf b}, we find that the coupling to the vector spin chirality modifies the dispersion without changing the topology of the bands for this choice of parameters. While the microscopic origin of interactions with vector and scalar spin chiralities which enter Eq.~\eqref{Ham} is fundamentally different, their roles for the resulting magnon dispersion are rather similar at the level of linear spin-wave theory.
Based on the obtained spin-wave spectra and Berry curvature calculations, we present in Fig.\ref{TOM_and_phase}{\bf c} the complete topological phase diagram as a function of the model parameters entering the  Hamiltonian. 
Sampling the nearest-neighbor coupling $J_2$, the DMI strength, and the magnitude of the $B$-field, we identify eight non-trivial phases 
in addition to an unstable ferromagnetic state. These phases come in pairs with an opposite overall sign in the set of Chern numbers.

\vspace{0.5cm}
\noindent
{\bf Imprinting orbital magnetism by magnons.}
To uncover the role of magnons in giving rise to orbital magnetism of the electrons through SSC, we evaluate the average value of the SSC that a given magnon carries, and translate it into the topological orbital moment of the magnon via the SSC-mediated orbital electron-magnon coupling. We refer to this quantity as
the  local TOM of the $n$-th magnon branch and access it according to $L_{n\mathbf k}^\mathrm{TOM}=\kappa^\mathrm{TO} \braket{\Psi_{n\mathbf k}|\chi(\mathbf k)|\Psi_{n\mathbf k}}$. Fig.~\ref{fig2}{\bf a--c} illustrates the value of the local TOM of the magnon branches as represented by the line thickness. While either finite DMI or $B$-field are necessary to activate the local TOM, the $\Gamma$ point typically hosts the minima and maxima of $L_{n\mathbf k}^\mathrm{TOM}$ in our model. Specifically, the local TOM of the lowest spin-wave branch reaches its global minimum at $\Gamma$ whereas the higher magnon bands carry the maximal values as they correspond to precessional modes with an innately larger  SSC. 
Clearly, the complex interplay between DMI and the orbital Zeeman coupling modifies not only the magnon topology but imprints also on the local TOM. In particular, the ordering of the states with positive and negative sign of $L_{n\mathbf k}^\mathrm{TOM}$ is inverted during the topological phase transition, which directly links the the nature of electronic orbital magnetism with non-trivial topology of magnonic bands.

Since the local orbital moment carried by magnons depends strongly on the band and position in the Brillouin zone, the effect of finite temperature which results in the excitation of magnons with finite energy, can give rise to a net magnon-mediated electronic orbital magnetization. To show this, we introduce a finite temperature $T$ in our spin system, and calculate the orbital response of the electronic bath. In Fig.~\ref{fig2}{\bf d--g} we analyze the sum of the local TOM weighted by the occupation number of each spin-wave branch at a given temperature, i.e., $\ell(\mathbf k)=\sum_n L_{n\mathbf k}^\mathrm{TOM} n_\mathrm{B}(\epsilon_{n\mathbf k})$. Here, the magnons follow the Bose distribution function $n_\mathrm{B}(\epsilon)=[\exp{(\beta\epsilon)}-1]^{-1}$ with $\beta=1/k_\mathrm{B}T$. Depending on $T$, the number of excited magnons is different in each branch, which leads to a non-trivial distribution of  $\ell(\mathbf k)$ in momentum space, as shown in Fig.~\ref{fig2}{\bf d--g} for the model with finite $B$-field but zero DMI. 
At low $T$, Fig.~\ref{fig2}{\bf d},
only the $\Gamma$-point magnons from the first branch can be excited, leading only to small local contributions around the BZ center. As the temperature is increased, 
all spin-wave states from the first branch are excited such that $\ell(\mathbf k)$ peaks in the $\mathrm M$ point with moderate magnitude as shown in Fig.~\ref{fig2}{\bf e}. If additionally magnons from the higher branches contribute at elevated temperatures, the maximum of $\ell(\mathbf k)$ occurs at the $\Gamma$ point, where the local TOM of the corresponding magnon states is the largest.

The overall TOM of the spin-wave system at given $T$ can be then obtained as:
\begin{equation}
\begin{split}
&\braket{L^\mathrm{TOM}}_T = \int\displaylimits_{\rm BZ} \ell(\mathbf k) \, d\mathbf k = \sum_n \int\displaylimits_{\rm BZ} n_\mathrm{B}(\epsilon_{n\mathbf k}) \, L_{n\mathbf k}^\mathrm{TOM}\, d\mathbf k,
\end{split}
\label{TOM}
\end{equation} 
where $\braket{L^\mathrm{TOM}}_T$ is the total TOM carried by thermally activated magnons per unit cell (see Supplementary Note 1)
Fig.~\ref{fig2}{\bf h--j} illustrates the $B,T$-dependence of the overall TOM 
for various DMI coupling strengths. On the one hand, as more magnons become available to carry the TOM, higher temperatures enhance the magnitude of $\braket{L^\mathrm{TOM}}_T$ in the spin-wave system. On the other hand, the roles of orbital Zeeman coupling and DMI are intertwined in generating TOM. 
For example, while TOM locally vanishes at zero DMI and $B$-field, a DMI with positive coupling strength generally counteracts the effect of the magnetic field on TOM if $\kappa^{\rm TO}$ is negative. 
For non-trivial choices of these parameters, however, Fig.~\ref{fig2}{\bf k} illustrates that at low $T$ the total TOM increases linearly, and, depending on the value of  $\kappa^{\rm TO}$, it can be sizeable. 

The total topological orbital moment emerges as a quantity which can be readily measured experimentally by referring to techniques which are sensitive to orbital magnetization in solids~\cite{go2018intrinsic,go2020orbital,li2019experimental,ding2020harnessing}. The sizeable magnitude of the effect that we predict not only lends itself to an unambiguous observation, but can also influence significantly the temperature dependence of the overall magnetization in a sample, providing thus an additional ``anomalous" orbital channel to the conventional mechanism of magnetization variation mediated by thermally excited magnons~\cite{bauer2012spin,boona2014spin,chumak2015magnon}. Given the much stronger  sensitivity of topological orbital magnetism to electronic structure changes, as compared to the spin, we suggest that the magnon-driven orbital magnetism can serve as a unique tool in tracking the electronic structure dynamics in various types of setups.   
As we also observe that the sign of $\braket{L^\mathrm{TOM}}_T$ correlates with the ordering of the topological spin-wave bands and their respective Chern numbers, we suggest to exploit the total topological orbital moment as an indicator of topological dynamics of magnons. 

\begin{figure}
\centering
\includegraphics[scale=0.265,trim=0 10 0 10]{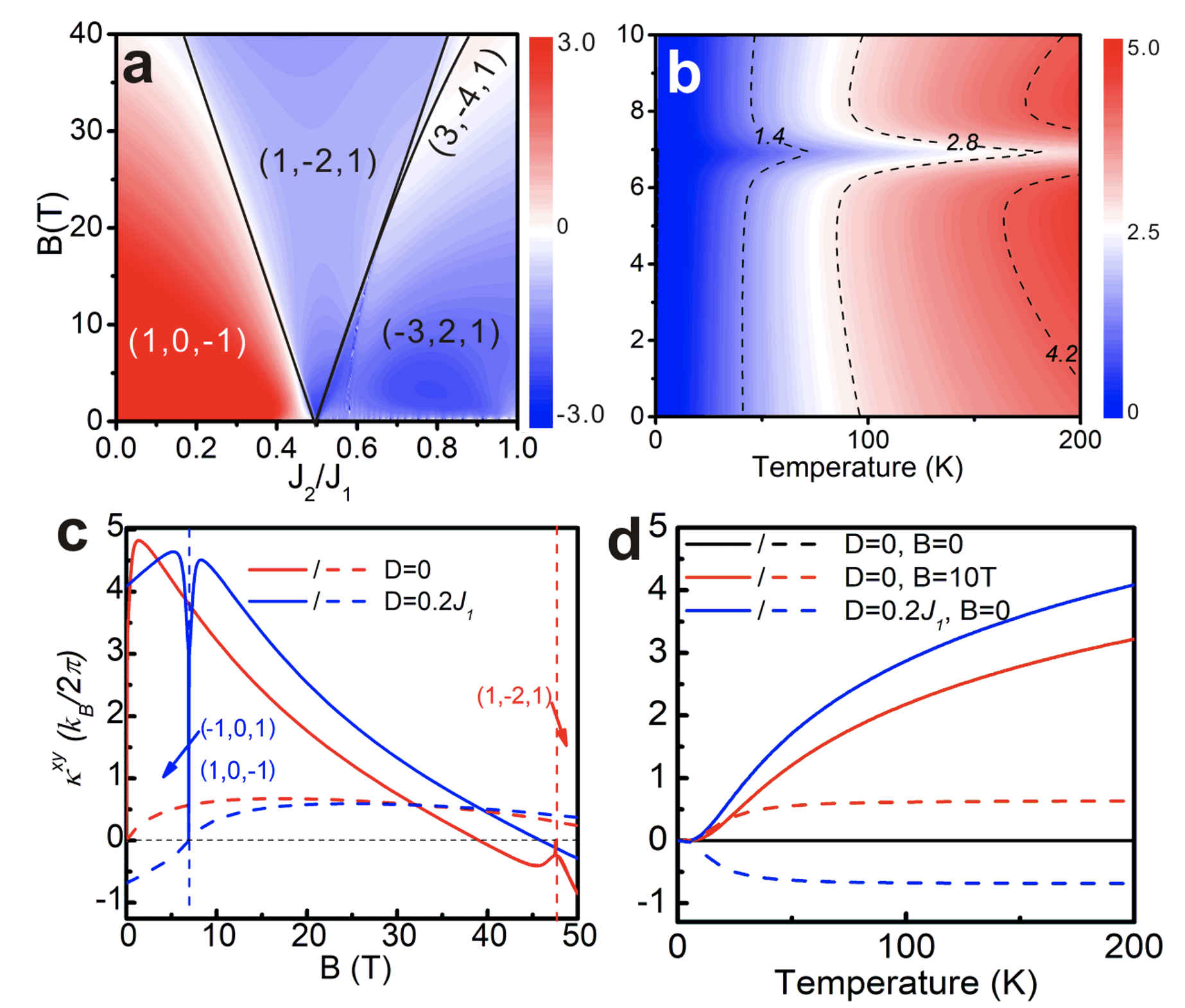}
\caption{{\bf Driving orbital currents by magnons: the orbital Nernst effect}. {\bf a}~Phase diagram of the orbital Nernst effect. Dependence of the orbital Nernst conductivity $\kappa_\mathrm{ONE}^{xy}$ on magnetic field $B$ and second nearest-neighbor Heisenberg coupling $J_2$ (in units of the nearest-neighbor Heisenberg coupling $J_1$) at $T=200$\,K and zero Dzyaloshinskii- Moriya interaction (DMI). Solid black lines are the boundaries between different topological phases characterized by the Chern numbers of the three magnon branches.  {\bf b}~$\kappa_\mathrm{ONE}^{xy}$ as a function of $B$ and temperature $T$ for the model with DMI strength of  $0.2J_1$. {\bf c,d}~Comparison of the $\kappa_\mathrm{ONE}^{xy}$ (solid lines) and magnon Nernst conductivity $\kappa_\mathrm{N}^{xy}$ (dashed lines). {\bf c}~  $\kappa_\mathrm{ONE}^{xy}$ and $\kappa_\mathrm{N}^{xy}$ as a function of $B$ for the model at 200~K with DMI strength of 0 (red) and $0.2J_1$ (blue). The different topological phases are distinguished with a thin vertical line. {\bf d}~ $\kappa_\mathrm{ONE}^{xy}$ and $\kappa_\mathrm{N}^{xy}$ as a function of $T$ for different strengths of the DMI and $B$.}
\label{fig3}
\end{figure}

\vspace{0.5cm}
\noindent
{\bf Driving orbital currents by magnons.}
Answering the first question posed in the introduction, our
analysis demonstrates that a finite TOM, stemming from orbital electronic currents,  can be
triggered by thermally activated magnons. 

This observation suggests that TOM is intimately linked to thermal spin transport which is mediated by the coupling of the SSC to the bath of electrons in the system. As a consequence, the well-known magnon Nernst effect acquires a novel and fundamentally distinct contribution that we coin the orbital Nernst effect of magnons, which is illustrated in Fig.~\ref{fig1}{\bf a}. The phenomenon of orbital Nernst effect relates spatial temperature gradients to the emergence of topological orbital currents via $j_x^\mathrm{TOM}=\kappa^{xy}_\mathrm{ONE}(\nabla T)_y$, where $\kappa_\mathrm{ONE}^{xy}$ stands for the topological orbital Nernst conductivity, which within the semiclassical theory reads
\begin{equation}
\begin{split}
    \kappa_\mathrm{ONE}^{xy}=-\frac{k_\mathrm{B}}{4\pi^2\mu_\mathrm{B} }\sum_n\int\displaylimits_{\rm BZ} c_1(n_\mathrm{B}(\epsilon_{n\mathbf k}))\,\Omega_{n\mathbf k}^{xy}
    L_{n\mathbf k}^\mathrm{TOM} \, d\mathbf k,
\end{split}
\label{eq:tone}
\end{equation}
where 
$c_1(\tau)=\int_0^{\tau}\ln[(1+t)/t]dt=(1+\tau)\ln(1+\tau)-\tau\ln \tau$. In essence, the latter relation quantifies the fundamental mechanism behind a magnon $-$ which develops a transverse velocity proportional to the Berry curvature in an applied temperature gradient $-$ ``dragging" with it the electronic orbital angular momentum which is generated by non-zero SSC inherent to the magnon. 
In contrast to the usual spin Nernst effect of magnons~\cite{matsumoto2011theoretical,cheng2016spin,kovalev2016spin}, the conductivity in Eq.~\eqref{eq:tone} characterizing the orbital Nernst effect depends explicitly on the local TOM of the magnon branches (see Supplementary Note 1).

Answering the second fundamental question posed in the introduction, below we reveal the existence of this effect by explicit calculation.
 In Fig.~\ref{fig3} we summarize the non-trivial dependence of the orbital Nernst effect on $T$ and on the model parameters, as well as its correlation with the topology of the magnon bands. Although the orbital Nernst effect has a distinct microscopic origin in the orbital electron-magnon coupling, our prediction is that the corresponding conductivity can reach the order of $\mathrm{\pi}^{-1}k_\mathrm{B}$. If we assume a distance of $5$~\AA{} between two kagome layers, an orbital Nernst conductivity of $(2\mathrm{\pi})^{-1} k_\mathrm{B}$ is equivalent to the value $4.394\times 10^{-15}$~Jm$^{-1}$K$^{-1}$, or $66786$~$ \hbar \mathrm e^{-1}\mu$Acm$^{-1}$K$^{-1}$, which is comparable to the values known for the spin Nernst effect of magnons or spin Nernst effect of electrons~\cite{kovalev2016spin,cheng2016spin,mook2014magnon,mook2019thermal,geranton2015spin,long2016giant}. We emphasize that the magnitude of the effect can be further enhanced by proper electronic structure engineering of the topological orbital susceptibility, which in its nature does not rely on the presence of spin-orbit interaction in the system.
 This underlines the strong potential of the orbital Nernst effect for the realm of spincaloritronics and marks this effect as an entry point for ideas evolving around magnon-mediated orbitronics.

\begin{figure}
\centering
\includegraphics[scale=0.23,trim=0 10 0 10]{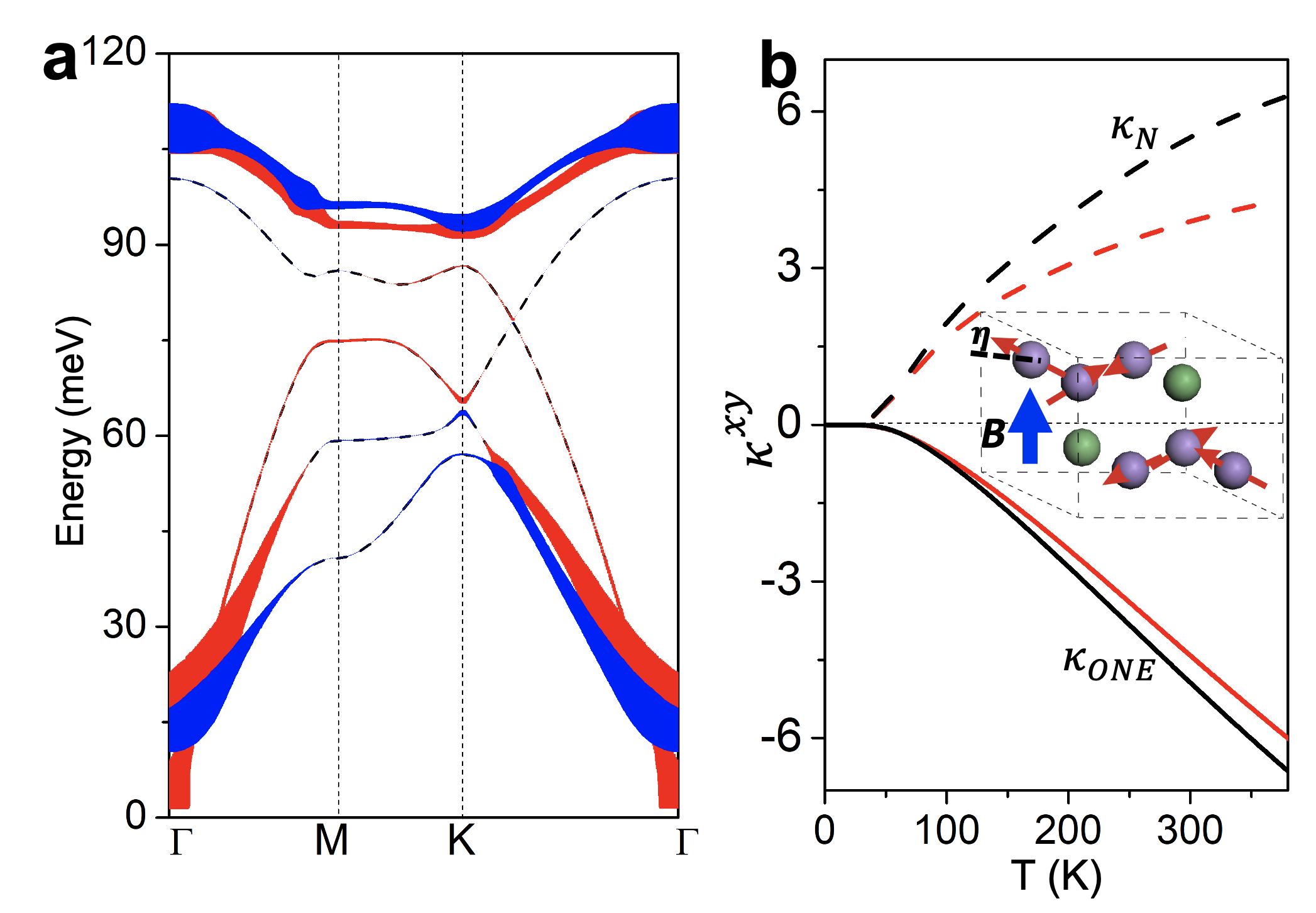}
\caption{{\bf Orbital Nernst effect in Mn$_3$Ge}. {\bf a}~Fat band analysis for the magnonic bands of Mn$_3$Ge with the canting angle $\eta=1^{\circ}$. Red and blue colors represent positive and negative sign of the local topological orbital moment (TOM) $L_{n\mathbf k}^\mathrm{TOM}$, respectively, and the line thickness denotes the corresponding magnitude. {\bf b}~Comparison between the magnon Nernst conductivity (dotted line) and orbital Nernst conductivity (solid line) as the function of temperature for  Mn$_3$Ge. Different color represents two different canting angles $\eta=0.4^{\circ}$ (red color) and $\eta=1^{\circ}$ (black color). The unit of Nernst conductivity in {
\bf b} is $10^3 \hbar e^{-1}\mu $A(cm)$^{-1}$K$^{-1}$. The schematic sketch of the magnetic structure of Mn$_3$Ge is shown in the inset.  }
\label{Mn3Ge}
\end{figure}

Our analysis,
which is supported by our calculation (see Fig.~\ref{fig3} and Supplementary Note 1), 
reveals that  both DMI and the coupling of external magnetic field to the SSC can generate a finite orbital Nernst conductivity. Comparing the two panels in more detail, we note that the sign of $\kappa_\mathrm{ONE}^{xy}$ is the same in topological phases for which the sets of Chern numbers differ by a global sign. This invariance stems from the product of the two microscopic quantities in Eq.~\eqref{eq:tone}, each of which changes its individual sign as the Chern numbers 
are reversed. Still, as exemplified in Fig.~\ref{fig3}{\bf a-c}, the orbital Nernst effect is characteristic to the non-trivial magnon topology of distinct phases.  
Close to topological phase transitions, the orbital Nernst effect changes abruptly and thus behaves rather differently compared to thermal Hall and magnon Nernst effects, see Fig.~\ref{fig3}{\bf c} and Supplementary Figure 11. As a consequence, the conductivity $\kappa_\mathrm{ONE}^{xy}$ can in principle reach very large values near the phase boundary. 
Since the orbital Nernst effect is absent without the $B$-field and DMI (see Supplementary Figure~9{\bf a}), the peak structure in Fig.~\ref{fig3}{\bf b,c} for a magnetic field of about $7$\,T can be understood as a result of the competition between the effects of orbital Zeeman coupling and DMI, 
which results in a strongly suppressed orbital Nernst effect.
On the other hand, Fig.~\ref{fig3}{\bf d} and Supplementary Figure~9{\bf b} reveal the qualitative difference in the temperature dependence of the orbital Nernst effect and conventional Nernst effect. The peculiar behavior of the orbital Nernst effect in response to an external magnetic field can be used to disentangle it from the magnon Nernst effect experimentally. Overall, the orbital Nernst effect presents a unique playground for generating orbital currents in magnonic systems, and we outline the prospects of this effect below.

\vspace{0.5cm}
\noindent
{\large{\bf Discussion}}\\
While in our work we consider ferromagnets on a kagome lattice, among material representatives of which one can name for example Cu(1-3,bdc)~\cite{chisnell2015topological} or Nd$_3$Sb$_3$Mg$_2$O$_{14}$~\cite{scheie2016effective}, the conclusions drawn from our analysis go well beyond this particular class of materials, and 
include for instance collinear or non-collinear states on a  hexagonal, pyrochlore, B20  and  Mn$_3$Ge quasi-kagome type of lattice~\cite{chen2018topological,hirschberger2015large,hanke2016role,grytsiuk2020topological,nayak2016large}, as well as their thin films. While in the latter classes the magnon drag of orbital momentum is non-vanishing, a precursor of prominent magnon-driven orbital phenomena is a large topological orbital susceptibility $\kappa^\mathrm{TO}$ in a given material of the order of that exhibited, e.g., by MnGe~\cite{grytsiuk2020topological}, Mn/Cu(111)~\cite{hanke2016role}, or Mn$_3$Ge~\cite{wimmer2019chirality}. The latter quantity can be estimated from microscopic calculations, as well as from experiment, as to first approximation $\kappa^\mathrm{TO}$ is given by the orbital susceptibility of the system~\cite{lux2018engineering}. 

To show this explicitly, 
we extract the rough magnitude of $\kappa^\mathrm{TO}$ from existing calculations
and specifically consider the case of Mn$_3$Ge~\cite{wimmer2019chirality}, exhibiting an almost coplanar non-collinear arrangement of spins.
By using the exchange parameters used to fit the experimental magnonic spectra, and taking into account a small canting of spins in Mn$_3$Ge in an external magnetic field (see Supplementary Note 2 and Supplementary Figure 12-17 for a detailed discussion), in Fig.~\ref{Mn3Ge} we provide the orbital analysis of the bands, and  the estimates for the magnon, $\kappa_\mathrm{N}^{xy}$, and orbital, $\kappa_\mathrm{ONE}^{xy}$, Nernst conductivities in this material as a function of temperature. Our calculations show that in Mn$_3$Ge the magnitude of magnonic and orbital contributions to the transverse thermal currents is comparable and sizeable. As both contributions are opposite in sign, this potentially gives rise to a non-trivial dependence of the overall current of angular momentum on temperature, which can be accessed experimentally. This signifies the potential relevance of discussed here orbital effects for wide classes of diverse magnetic materials.


The uncovered mechanism of magnon-driven chirality accumulation  has far-reaching consequences for the transport properties of systems which exhibit such chirality. For example, it will result
in the generation of topological Hall or topological spin Hall effect of the underlying electronic bath~\cite{bruno2004topological,buhl2017topological,franz2014real,buhl2017topological}, 
which will contribute to the temperature dependence of the anomalous Hall conductivity even  in nominally collinear magnets~\cite{ishizuka2018spin}.
On the other hand, magnon-driven orbital magnetism 
brings the orbital angular momentum variable into the game of magnon-based spincaloritronics, which is conventionally associated with generation and transport of spin. Unleashing the orbital channel for the magnon-mediated effects poses a key question of the role of orbital magnetism for the temperature-dependent magnetization dynamics, however, it also opens a number of exciting possibilities for direct applications. For example, in analogy to the spin-orbit torques~\cite{miron2010current,garello2013symmetry}, we envisage that the flow of orbital angular momentum generated by magnons can be used to generate sizeable orbital accumulation and orbital torques on adjacent magnets, which can go either via the mechanism of direct injection of the orbital current into the ferromagnet, or might involve an intermediate conversion of the orbital current into the spin current with the magnitude of the converted spin current larger by far than that driven by the local spin Hall effect~\cite{ding2020harnessing,go2018intrinsic,go2020orbital}. 

Given the sensitivity of the orbital effects to the topology of magnonic bands and generally magnonic properties, we suggest that accessing the magnon-mediated dynamics of orbital properties can serve as a unique tool of tracking the topological dynamics of magnons. Moreover, our findings also point at an exciting possibility of exploiting properly engineered orbital injection for excitation of specific magnonic modes via the inverse orbital Nernst effect. As in topologically-complex materials the electronic topology is directly related to the orbital properties~\cite{niu2019mixed}, this link can be used for realizing hybrid non-trivial electron-magnon topologies. Overall, the uncovered here orbital electron-magnon coupling bares various prospects for integration of spinorbitronics schemes into magnonic setups and vice versa, which shall be explored in the future.  

\vspace{0.5cm}
\noindent
{\large{\bf Methods}}\\
\noindent
{\bf Calculation of electronic TOM.} For the calculation of the electronic structure and resulting TOM, we employ a tight-binding model on a two-dimensional kagome lattice. The Hamiltonian consists of hoppings and local exchange interactions,
\begin{eqnarray}
\mathcal{H}
=
t_1 \sum_{\langle i,j \rangle} c_i^\dagger c_j + t_2 \sum_{ \llangle i,j \rrangle} c_i^\dagger c_j 
+
J \sum_{i} \hat{\mathbf{m}}_i \cdot \boldsymbol{\sigma},
\end{eqnarray}
where $i$ and $j$ are site indices, $\langle \cdots \rangle$ and $\llangle \cdots \rrangle$ indicate first and second nearest neighbor pairs, respectively, and $\hat{\mathbf{m}}_i$ is the direction of the local magnetic moment at site $i$. The first and second nearest hopping amplitudes are chosen as $t_1 = 1.0\ \mathrm{eV}$ and $t_2 = 0.15\ \mathrm{eV}$, respectively, and strength of the exchange interaction is set to $J=1.7\ \mathrm{eV}$. For three basis atoms in the unit cell, namely A, B, and C, the directions of the local magnetic moments are parametrized by $\hat{\mathbf{m}}_i = (\sin\theta\cos\phi_i, \sin\theta\sin\phi_i, \cos\theta)$. The azimuthal angles $\phi_i$ are assumed to be chirally ordered, i.e., $\phi_{i}=\phi_0$ for $i\in \mathrm{A}$, $\phi_{i}=\phi_0 + 2\pi/3$ for $i\in \mathrm{B}$, and $\phi_{i}=\phi_0 + 4\pi/3$ for $i\in \mathrm{C}$. For the fitting of the TOM in Fig.~\ref{TOM_and_phase}{\bf b}, we assume
\begin{eqnarray}
L_z^\mathrm{TOM} (\theta)
=
\kappa^\mathrm{TO}
\hat{\mathbf{m}}_A \cdot (\hat{\mathbf{m}}_B\times \hat{\mathbf{m}}_C)
=
\frac{3\sqrt{3}}{2}\kappa^\mathrm{TO} \cos\theta\sin^2\theta.
\nonumber
\\
\end{eqnarray}
To extract $\kappa^\mathrm{TO}$ near $\theta = 0$ (inset of Fig.~\ref{TOM_and_phase}{\bf b}), we use
\begin{eqnarray}
\kappa^\mathrm{TO}
=
\frac{2}{3\sqrt{3}} \left. \frac{d^2L_z^\mathrm{TOM}}{d\theta^2} \right |_{\theta=0},
\end{eqnarray}
where the second derivative is evaluated by a finite difference method.

\noindent
{\bf Linear spin wave theory.}
 Linear spin-wave theory~\cite{toth2015linear,mook2014magnon} is used to obtain the eigenvalues and eigenvectors of the above Hamiltonian, which we reformulate first in terms of bosonic ladder operators $a_i$ and $a_i^\dagger$ via the Holstein-Primakoff transformation~\cite{holstein1940field}.  In the resulting spin-wave Hamiltonian, we keep only terms that are quadratic in the ladder operators. This approximation has been used before to treat the effect of chirality~\cite{katsura2010theory}.
 Within linear theory, the SSC $\chi_{ijk}$, coupling directly to the magnetic field in Eq.~\eqref{Ham}, can be expressed as~\cite{katsura2010theory}:
\begin{equation}
    \chi_{ijk} =  {\frac{\mathrm i}{S}}\,(a_i^{\dag}a_j-a_ia_j^{\dag}+a_j^{\dag}a_k-a_ja_k^{\dag}+a_k^{\dag}a_i-a_ka_i^{\dag}) \, .
    \label{eq:2}
\end{equation}
To map from real to momentum space, we perform a Fourier transform of the bosonic ladder operators, which leads to the Hamiltonian matrix $H(\mathbf k)$ at the spin-wave vector $\mathbf k=(k_x,k_y)$,
  which is diagonalized to obtain the eigenvectors and the energy spectrum of the spin waves. We address the topological character of the magnonic bands by computing the Chern number $C_n$, given by 
$C_n=\frac{1}{2\pi}\int\Omega_{n\mathbf k}^{xy} \,d\mathbf k$, where the integral is performed over the Brillouin zone (BZ), and $\Omega_{n\mathbf k}^{xy}$ represents 
the magnon Berry curvature of the $n$th spin-wave branch:
\begin{equation}
 \Omega_{n\mathbf k}^{xy}=-2\,\mathrm{Im}\sum_{m\neq n}
 \frac{\braket{\Psi_{n\mathbf{k}}|\frac{\partial H(\mathbf{k})}{\partial k_x}|\Psi_{m\mathbf{k}}}\braket{\Psi_{m\mathbf{k}}|\frac{\partial H(\mathbf{k})}{\partial k_y}|\Psi_{n\mathbf{k}}}}{(\epsilon_{n\mathbf{k}}-\epsilon_{m\mathbf{k}})^2} \, ,
 \label{curvature}
 \end{equation}
where $|\Psi_{n\mathbf k}\rangle$ is an eigenstate of the spin-wave Hamiltonian with the energy $\epsilon_{n\mathbf k}$ (see Supplementary methods)
.

\vspace{0.5cm}
\noindent
{\large{\bf Data availability}}\\
The  data that support the findings of this study are available from the corresponding
author upon reasonable request.

\vspace{0.5cm}
\noindent
{\large{\bf Code availability}}\\
The code  of this work are available from the
corresponding authors on request.

\vspace{0.5cm}
\def\bibsection{\noindent{\large{\bf References}}}

\vspace{0.5cm}
\noindent
{\large{\bf Acknowledgements}}\\
We acknowledge fruitful discussions with Marjana Le\v{z}ai\'c and  Olena Gomonay. L.-C.~Zhang acknowledges support from China Scholarship Council (CSC) (No. [2016]3100).  This work was funded by the Deutsche Forschungsgemeinschaft (DFG, German Research Foundation) $-$ TRR 173 $-$ 268565370 (project A11), TRR 288 $-$ 422213477 (project B06), and SPP 2137 ``Skyrmionics" (project  MO  1731/7-1). We gratefully acknowledge financial support from the European Research Council (ERC) under the European Union's Horizon 2020 research and innovation program (Grant No. 856538, project "3D MAGiC").
We also gratefully acknowledge the J\"ulich Supercomputing Centre and RWTH Aachen University for providing computational resources under project jiff40.

\vspace{0.5cm}
\noindent
{\large{\bf Author contributions}}\\
F. R. L. and Y. M. conceived the concept. F. R. L., D. G. and Y. M. designed the research.  L.-C. Z. performed the magnonic part of the calculations and analysis, D. G. performed the calculations of the electronic properties. L.-C. Z., J.-P. H., D. G. and Y. M. wrote the manuscript. J.-P. H., S. G., P. M. B. and S. B. discussed the results and contributed to revising the manuscript with the rest of the authors.

\vspace{0.5cm}
\noindent
{\large{\bf Competing interests}}\\
The authors declare no competing interests. 
\end{document}